\documentclass[twocolumn, trackchanges]{aastex7}
\usepackage{xcolor}
\usepackage{amsmath}
\usepackage{longtable}
\usepackage{comment}
\usepackage{soul}

\begin{document}

\title{$r$-process Abundance Dispersion in the Globular Cluster M5 using Keck Archival Data}

\author[0009-0000-0107-0244]{Pranav Nalamwar}
\affiliation{Department of Physics and Astronomy, University of Notre Dame, \\
225 Nieuwland Science Hall, Notre Dame, IN 46556, USA}
\email{pnalamwa@nd.edu}

\author[0000-0001-6196-5162]{Evan N.\ Kirby}
\affiliation{Department of Physics and Astronomy, University of Notre Dame, \\
225 Nieuwland Science Hall, Notre Dame, IN 46556, USA}
\email{ekirby@nd.edu}

\author[0009-0001-0983-623X]{Alice Cai}
\affiliation{Center for Interdisciplinary Exploration and Research in Astrophysics (CIERA), Northwestern University, \\
1800 Sherman Ave 8th Floor, Evanston, IL 60201, USA}
\email{AliceCai2029@u.northwestern.edu}

\begin{abstract}

We studied $28$ RGB stars in the mildly metal-rich globular cluster M5 ([Fe/H] $= -1.29$) using archival high-resolution spectra from the Keck Observatory archive (KOA) to better understand the $r$-process in globular clusters. Previous studies (M15, M92, and NGC 2298) have shown $r$-process dispersion in varying amounts, hinting at the source of the $r$-process in those clusters. We extend these dispersion studies to the more metal-rich cluster M5 by studying the rare-earth peak, specifically the elements Ba, Nd, and Eu.  We separately analyze the different stellar generations, as traced by the abundance of Na and O\@. Based on the Nd and Eu abundances, we report a tenuous detection of $r$-process dispersion that is dependent on the generation and element. Based on a log-likelihood dispersion study accounting for measurement errors, Nd has an intrinsic first generation abundance spread of $\sigma_{1G}(\text{Nd}) = 0.15_{-0.07}^{+0.10}$ and an $2\sigma$ upper limit on the second generation spread of $\sigma_{2G}(\text{Nd}) < 0.28$. The upper limits on the Eu intrinsic spread are $\sigma_{1G}(\text{Eu}) < 0.34$ and $\sigma_{2G}(\text{Eu}) < 0.16$. A potential dispersion implies the cluster gas was inhomogeneously polluted, either due to an event concurrent with the formation of the cluster or due to clouds of disparate composition that coalesced to form the cluster.
\end{abstract}



\section{Introduction} \label{sec:intro} 

The rapid neutron capture process, or $r$-process, is the formation mechanism responsible for the heaviest elements of the universe such as lanthanum, gold, and uranium \citep{Burbidge1957}. Supernovae and neutron star mergers (NSMs) have been thoroughly studied as the sites for the main $r$-process (mass number $A \geq 140$), namely due to the high flux of neutrons present in these scenarios \citep{Cowan2021}. While supernovae were once thought to be capable of producing the main $r$-process, realistic 3D supernovae simulations with neutrino transport suggest the electron fraction, or proton-to-baryon ratio, is too high
to produce a heavy $r$-process due to the proliferation of weak reactions \citep{Burrows2020,Wanajo2023}. Even though the observation of a kilonova from the gravitational wave event GW170817 conclusively proved neutron star mergers (NSMs) to be a site of the $r$-process \citep{Abbot2017, Watson2019}, their rarity and long formation time call into question whether they can explain the solar $r$-process pattern or patterns found in other metal-poor stars \citep{Skuladottir19}. 

Other sites include exotic supernovae, such as collapsars and magneto-rotational supernovae, along with other stellar explosions like magnetar giant flares \citep{Cowan2021, Reichert2023, Patel2025}. These sites, while rare, occur on shorter timescales (on the order of tens of Myr) and thus are considered in the literature as ``prompt'' sources. On the other hand, while NSMs are also rare, they are considered ``long'' timescale events as the infall time for two neutron stars to merge is, on average, on the order of 1 Gyr. Additionally, an admixture of contributions from several sites have been postulated to explain the variety of $r$-process abundance patterns in metal-poor stars \citep{Wasserburg2000, CJHansen2014} .

Several stellar and galactic systems have been studied to investigate the site of the $r$-process, from Milky Way disk studies (e.g., the declining trend of [Eu/Fe] vs.\ [Fe/H], indicating at least one prompt r-process site, \citealt{Cote2019}) to abundance studies in dwarf galaxies \citep{Naidu2022}. Dwarf galaxies, especially the ultra-faint dwarf galaxies (UFDs), have low-stellar-mass populations that experienced few rare $r$-process-producing events during formation. Any rapid neutron-capture elements found, such as europium, would be from one or a few polluting events, thereby constraining the $r$-process site. However, even the $r$-process enhanced UFD Reticulum II is at a distance of $\approx 30$ kpc \citep{Bechtol2015,Koposov2015,Ji2016}, limiting high-resolution studies for detailed abundances to the brightest stars, which are few in number given the low mass of dwarf galaxies.

Globular clusters (GCs), on the other hand, offer an alternative approach to studying the $r$-process. Globular clusters are clusters of stars ranging from $10^4 - 10^6$ \(M_\odot\) in present-day mass and are found within the metal-rich dominated galactic bulge and the metal-poor dominated galactic halo. They were thought to have formed from a single nebula as metallicity is uniform throughout the cluster (with some notable exceptions, like $\omega$ Centauri). However, essentially all spectroscopic studies of these objects have shown anti-correlations between the abundances of certain light elements such as sodium and oxygen \citep{Bragaglia2017, Gratton2019}, leading to the multiple population/generation phenomenon. This light-element dispersion led \cite{Carretta2009} to define GCs as clusters exhibiting light-element dispersions, a stark contrast from the original assumptions. 

$r$-process dispersion, however, may not be ubiquitous. To date, only 3 clusters, M15 \citep{Sneden2000, Cabrera-Garcia2024}, M92 \citep{Kirby2023}, and NGC 2298 \citep{Bandyopadhyay2025} have proven variations in $r$-process elements, such as the lanthanides. In addition, NGC 1261 shows signs of $r$-process dispersion, which requires followup studies to confirm \citep{Koch-Hansen2021}. Coupling the nature of globular cluster formation \citep{Krause2020} with the existence of $r$-process dispersion may provide an avenue to ascertain the possible $r$-process source(s). Given the dispersion in M92, especially within the first generation, \cite{Kirby2023} suggested M92 experienced a prompt, prolific $r$-process event prior to or during the formation of the first generation. A similar argument was made by \cite{Cabrera-Garcia2024} for M15. Because M15, M92, and NGC 228 are all metal-poor clusters, it is interesting to explore slightly more metal-rich GCs ($-2 \leq {\rm [Fe/H]} \leq -1$) to determine whether the origin of the $r$-process is ubiquitous across GC environments. \cite{Roederer2011} highlighted several other clusters with potential for $r$-process dispersion, namely NGC~5904 (M5), NGC~3201, NGC~5272 (M3), and NGC~6205 (M13). Other than M15 and M92, they indicated M5 has the highest probability of displaying $r$-process dispersion out of the northern clusters. %Keck Observatory suits such studies in the northern hemisphere.

To this end, we studied the Keck Observatory Archive (KOA) High-Resolution Echelle Spectrometer (HIRES) spectra for the massive, intermediate metallicity (${\rm [Fe/H]} = -1.29$) globular cluster M5. We quantify the dispersions present in the main $r$-process elements Nd and Eu within our sample, along with a comparison between stellar generations.

\section{Methodology}

\subsection{Keck Observatory Archival Data}
\label{sec:Observations}

The KOA includes HIRES spectra for asymptotic giant branch (AGB) and red giant branch (RGB) stars in M5. To reduce stellar modeling uncertainties, we only analyzed RGBs. We chose 28 RGBs that were readily associated with previous photometric surveys \citep{Arp1955, Stetson2019} and had previous literature confirming their stellar classification \citep{Ivans2001,Ramirez2003, Sandquist2004, Lai2011}.

HIRES was upgraded in 2004 to increase its optical efficiency and spectral range. $22$ stars within this sample were observed before this upgrade, resulting in the limited spectral range $5200 \lesssim \lambda \lesssim 7700 $ \AA\@. The other $6$ stars were observed after the upgrade, so they have wider spectra with a range of $3800 \lesssim \lambda \lesssim 8000$ \AA\@. Table \ref{tab:stellar_data} presents the star list with the appropriate stellar parameters, spectral information, and the $\log \epsilon$ abundances for the elements within our study.   

\begin{deluxetable*}{lccccccccccc}
\tablecolumns{11}
\tablewidth{\textwidth}  % Force table to fit within text width
\tabletypesize{\scriptsize} % Reduce font size to fit long rows
\tablecaption{Star info and derived abundances for M5 KOA sample. Note that all stars in this study are RGB stars with an [Fe/H] = $-1.29$ and an [$\alpha$/Fe] = $+0.4$. The S/N is computed in the region $\lambda = 7100 ~ - ~ 7120$ \AA. \label{tab:stellar_data}}
\tablehead{
\colhead{KOA Star} & 
\colhead{$T_{\text{eff}}$} & 
\colhead{$\log g$} & 
\colhead{$\xi$} & 
\colhead{[O/Fe]} & 
\colhead{[Na/Fe]} & 
\colhead{[Ba/Fe]} & 
\colhead{[Nd/Fe]} & 
\colhead{[Eu/Fe]} & 
\colhead{S/N} &
\colhead{P.I.} & 
\colhead{Obs. Yr}}
\startdata
I-14 & 4412.46 $\pm$ 83.0 & 1.11 $\pm$ 0.04 & 1.87 $\pm$ 0.03 & 0.17 $\pm$ 0.24 & 0.44 $\pm$ 0.1 & 0.29 $\pm$ 0.2 & - & 0.77 $\pm$ 0.13 & 61.58 & R.Kraft & 1995 \\
II-59 & 4537.12 $\pm$ 83.0 & 1.32 $\pm$ 0.03 & 1.83 $\pm$ 0.04 & - & 0.57 $\pm$ 0.1 & 0.06 $\pm$ 0.13 & - & 0.67 $\pm$ 0.13 & 38.48 & R.Kraft & 1995 \\
I-50 & 4697.13 $\pm$ 83.0 & 1.62 $\pm$ 0.03 & 1.76 $\pm$ 0.05 & - & 0.53 $\pm$ 0.12 & 0.11 $\pm$ 0.1 & - & 0.72 $\pm$ 0.11 & 73.69 & R.Kraft & 1995 \\
IV-34 & 4435.61 $\pm$ 83.0 & 1.14 $\pm$ 0.04 & 1.87 $\pm$ 0.04 & 0.45 $\pm$ 0.06 & 0.45 $\pm$ 0.11 & 0.14 $\pm$ 0.12 & - & 0.75 $\pm$ 0.11 & 78.26 & R.Kraft & 1995 \\
I-58 & 4516.72 $\pm$ 83.0 & 1.28 $\pm$ 0.03 & 1.83 $\pm$ 0.04 & 0.38 $\pm$ 0.08 & 0.44 $\pm$ 0.11 & 0.14 $\pm$ 0.11 & - & 0.76 $\pm$ 0.11 & 72.8 & R.Kraft & 1995 \\
IV-47 & 4161.17 $\pm$ 83.0 & 0.68 $\pm$ 0.04 & 1.97 $\pm$ 0.02 & 0.56 $\pm$ 0.07 & 0.52 $\pm$ 0.1 & 0.05 $\pm$ 0.14 & - & 0.6 $\pm$ 0.13 & 51.23 & M.Rich & 1998 \\
II-85 & 4101.88 $\pm$ 83.0 & 0.61 $\pm$ 0.04 & 1.99 $\pm$ 0.02 & 0.74 $\pm$ 0.08 & 0.5 $\pm$ 0.1 & -0.4 $\pm$ 0.14 & - & 0.3 $\pm$ 0.13 & 52.08 & M.Rich & 1998 \\
III-66 & 4765.6 $\pm$ 83.0 & 1.78 $\pm$ 0.03 & 1.72 $\pm$ 0.05 & - & 0.66 $\pm$ 0.15 & -0.18 $\pm$ 0.14 & 0.73 $\pm$ 0.14 & - & 25.91 & Graeme.Smith & 1999 \\
IV-3 & 4992.8 $\pm$ 83.0 & 2.25 $\pm$ 0.03 & 1.61 $\pm$ 0.07 & - & 0.18 $\pm$ 0.16 & 0.01 $\pm$ 0.14 & 0.01 $\pm$ 0.14 & 0.61 $\pm$ 0.15 & 26.61 & Graeme.Smith & 1999 \\
I-43 & 4735.33 $\pm$ 83.0 & 1.71 $\pm$ 0.03 & 1.74 $\pm$ 0.05 & 0.64 $\pm$ 0.09 & 0.1 $\pm$ 0.1 & 0.05 $\pm$ 0.11 & 0.38 $\pm$ 0.09 & 0.66 $\pm$ 0.11 & 71.24 & Graeme.Smith & 1999 \\
I-21 & 4733.78 $\pm$ 83.0 & 1.69 $\pm$ 0.03 & 1.74 $\pm$ 0.05 & 0.65 $\pm$ 0.09 & 0.56 $\pm$ 0.11 & 0.08 $\pm$ 0.1 & 0.36 $\pm$ 0.09 & 0.61 $\pm$ 0.1 & 70.48 & Graeme.Smith & 1999 \\
I-65 & 5003.28 $\pm$ 83.0 & 2.32 $\pm$ 0.03 & 1.6 $\pm$ 0.07 & - & 0.34 $\pm$ 0.11 & 0.0 $\pm$ 0.13 & 0.23 $\pm$ 0.09 & 0.73 $\pm$ 0.12 & 67.12 & Graeme.Smith & 2000 \\
IV-24 & 5003.44 $\pm$ 83.0 & 2.31 $\pm$ 0.03 & 1.6 $\pm$ 0.07 & 0.98 $\pm$ 0.11 & 0.28 $\pm$ 0.09 & 0.01 $\pm$ 0.15 & 0.33 $\pm$ 0.07 & 0.51 $\pm$ 0.13 & 67.84 & Graeme.Smith & 2000 \\
IV-74 & 4601.63 $\pm$ 83.0 & 1.44 $\pm$ 0.03 & 1.8 $\pm$ 0.04 & 0.54 $\pm$ 0.06 & 0.58 $\pm$ 0.11 & 0.16 $\pm$ 0.12 & 0.45 $\pm$ 0.06 & 0.72 $\pm$ 0.12 & 63.05 & Graeme.Smith & 2000 \\
II-84 & 4983.74 $\pm$ 83.0 & 2.27 $\pm$ 0.03 & 1.61 $\pm$ 0.07 & 1.31 $\pm$ 0.09 & 0.54 $\pm$ 0.12 & -0.16 $\pm$ 0.14 & 0.27 $\pm$ 0.08 & 0.54 $\pm$ 0.13 & 40.48 & Graeme.Smith & 2000 \\
I-39 & 4464.88 $\pm$ 83.0 & 1.19 $\pm$ 0.04 & 1.86 $\pm$ 0.04 & 0.58 $\pm$ 0.11 & 0.61 $\pm$ 0.11 & 0.14 $\pm$ 0.11 & 0.36 $\pm$ 0.07 & 0.74 $\pm$ 0.11 & 44.27 & Graeme.Smith & 2000 \\
I-4 & 4554.39 $\pm$ 83.0 & 1.36 $\pm$ 0.03 & 1.82 $\pm$ 0.04 & 0.86 $\pm$ 0.09 & 0.15 $\pm$ 0.09 & 0.05 $\pm$ 0.11 & 0.41 $\pm$ 0.07 & 0.7 $\pm$ 0.11 & 88.98 & Graeme.Smith & 2000 \\
II-16 & 4972.25 $\pm$ 83.0 & 2.25 $\pm$ 0.03 & 1.61 $\pm$ 0.07 & 0.72 $\pm$ 0.08 & 0.08 $\pm$ 0.08 & 0.02 $\pm$ 0.11 & 0.33 $\pm$ 0.07 & 0.47 $\pm$ 0.12 & 74.77 & Graeme.Smith & 2000 \\
III-41 & 4965.99 $\pm$ 83.0 & 2.25 $\pm$ 0.03 & 1.61 $\pm$ 0.07 & - & 0.33 $\pm$ 0.11 & -0.1 $\pm$ 0.11 & 0.21 $\pm$ 0.1 & 0.16 $\pm$ 0.14 & 59.39 & Graeme.Smith & 2000 \\
I-25 & 4626.85 $\pm$ 83.0 & 1.49 $\pm$ 0.03 & 1.79 $\pm$ 0.05 & 0.52 $\pm$ 0.08 & 0.42 $\pm$ 0.1 & 0.08 $\pm$ 0.11 & 0.43 $\pm$ 0.09 & 0.69 $\pm$ 0.11 & 71.31 & Graeme.Smith & 2000 \\
II-17 & 4967.28 $\pm$ 83.0 & 2.28 $\pm$ 0.03 & 1.61 $\pm$ 0.07 & - & 0.11 $\pm$ 0.09 & 0.06 $\pm$ 0.11 & 0.27 $\pm$ 0.08 & 0.69 $\pm$ 0.14 & 63.3 & Graeme.Smith & 2000 \\
IV-81 & 4039.47 $\pm$ 83.0 & 0.5 $\pm$ 0.04 & 2.01 $\pm$ 0.02 & 0.62 $\pm$ 0.07 & 0.45 $\pm$ 0.1 & -0.11 $\pm$ 0.16 & 0.42 $\pm$ 0.1 & 0.56 $\pm$ 0.17 & 50.35 & J.Cohen & 2005 \\
R21 & 4177.57 $\pm$ 83.0 & 0.75 $\pm$ 0.04 & 1.96 $\pm$ 0.02 & 0.58 $\pm$ 0.08 & 0.23 $\pm$ 0.1 & -0.13 $\pm$ 0.16 & 0.47 $\pm$ 0.11 & 0.36 $\pm$ 0.15 & 74.48 & Bolte & 2007 \\
R9 & 4114.04 $\pm$ 83.0 & 0.6 $\pm$ 0.04 & 1.99 $\pm$ 0.02 & -0.09 $\pm$ 0.1 & 0.47 $\pm$ 0.11 & 0.05 $\pm$ 0.12 & 0.45 $\pm$ 0.1 & 0.61 $\pm$ 0.1 & 49.29 & Bolte & 2007 \\
S1056 & 4226.15 $\pm$ 83.0 & 0.82 $\pm$ 0.04 & 1.94 $\pm$ 0.03 & 0.64 $\pm$ 0.05 & 0.13 $\pm$ 0.1 & 0.23 $\pm$ 0.15 & 0.78 $\pm$ 0.16 & 0.71 $\pm$ 0.12 & 80.13 & J.Cohen & 2012 \\
S183 & 4251.58 $\pm$ 83.0 & 0.82 $\pm$ 0.04 & 1.94 $\pm$ 0.03 & 0.4 $\pm$ 0.05 & 0.54 $\pm$ 0.11 & 0.24 $\pm$ 0.15 & 0.69 $\pm$ 0.1 & 0.76 $\pm$ 0.15 & 58.15 & J.Cohen & 2012 \\
S1304 & 4241.47 $\pm$ 83.0 & 0.83 $\pm$ 0.04 & 1.94 $\pm$ 0.03 & 0.65 $\pm$ 0.06 & 0.21 $\pm$ 0.13 & 0.13 $\pm$ 0.15 & 0.64 $\pm$ 0.09 & 0.65 $\pm$ 0.11 & 74.2 & J.Cohen & 2012 \\
S1174 & 4353.68 $\pm$ 83.0 & 1.02 $\pm$ 0.04 & 1.89 $\pm$ 0.03 & 0.73 $\pm$ 0.07 & 0.11 $\pm$ 0.1 & 0.27 $\pm$ 0.16 & 0.63 $\pm$ 0.09 & 0.94 $\pm$ 0.13 & 95.11 & J.Cohen & 2012
\enddata
\end{deluxetable*}

\begin{figure*}
    \centering
    \includegraphics[width=\linewidth]{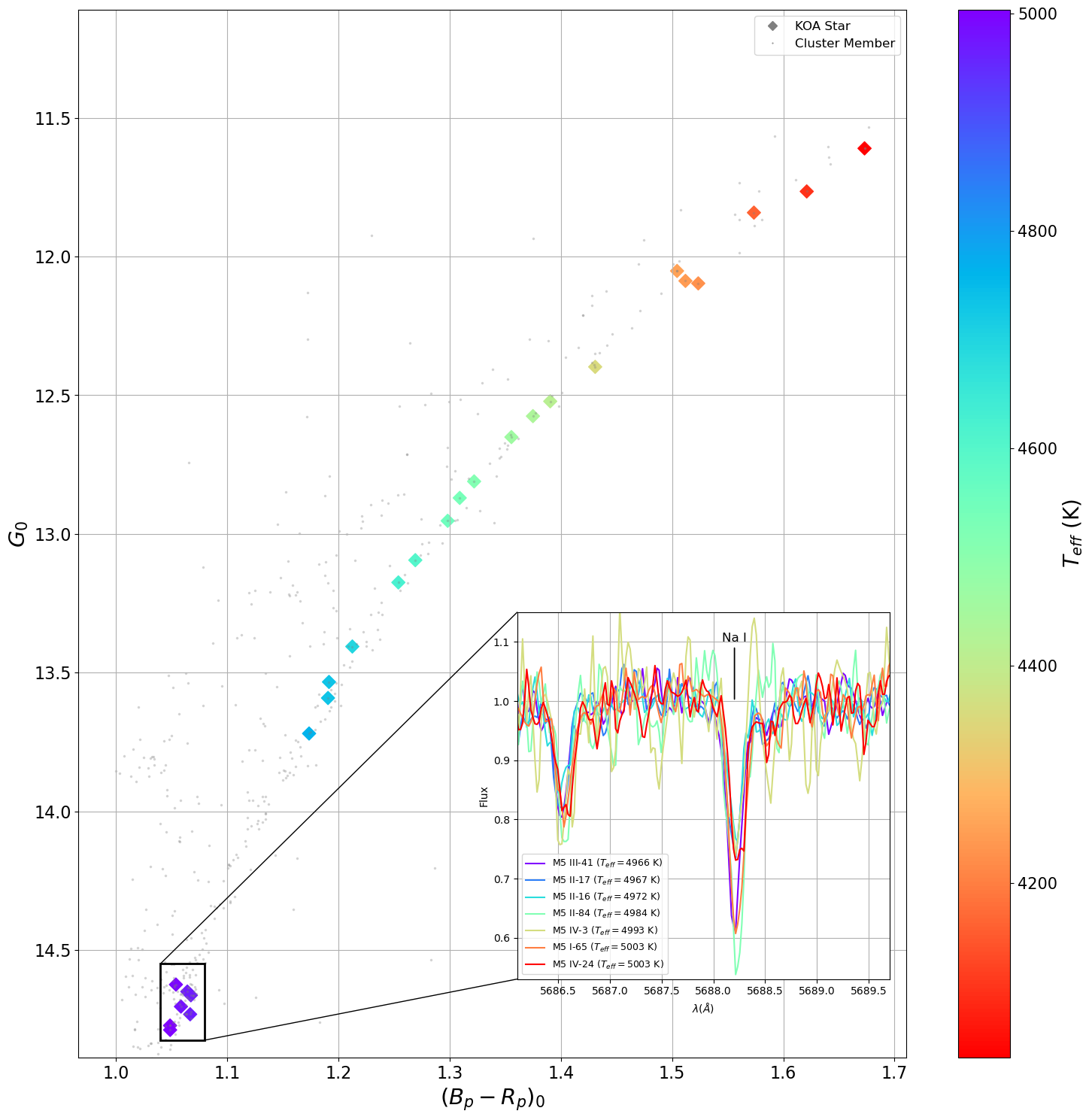}
    \caption{Color--magnitude diagram for the $28$ RGBs in our sample, color-coded by effective temperature. The inset shows the spectra of the $7$ hottest RGBs at the \ion{Na}{1} $\lambda \hspace{.5em} 5688$~\AA\ line.  (The colors for the spectra do not correspond to temperature and are only shown to differentiate the individual spectra.) The stars  have similar temperatures but different line strengths, demonstrating the well-known sodium abundance dispersion in M5.}
    \label{fig:CMD_spectra}
\end{figure*}

\subsection{Spectrum Preparation}
\label{sec: spectrum}

For each exposure's echelle order, the KOA stores raw and reduced spectra from HIRES\@. We used the reduced spectra, which are already one-dimensional, wavelength-calibrated, and sky-subtracted by the reduction pipeline {\fontfamily{qcr}\selectfont
MAKEE} \citep{Barlow2024}. Prior to equivalent width analysis, we continuum-normalized each reduced echelle order and coadded all normalized orders in the following manner.  

%. Should I explain what continuum normalization is? % I think that's probably unnecessary, the name is somewhat self-explanatory enough

For each star, we interpolated an {\fontfamily{qcr}\selectfont
ATLAS9} \citep{Castelli2003} model stellar atmosphere using stellar parameters derived in section \ref{sec: stellar_params} and the model atmosphere grid from \cite{Kirby2011}.
For the synthetic spectrum of each star, we used  {\fontfamily{qcr}\selectfont
MOOG}, which took as input a line list generated via the code {\fontfamily{qcr}\selectfont
LINEMAKE} \citep{Placco2021}. Our line list for  {\fontfamily{qcr}\selectfont
MOOG} comprises atomic, molecular, and hyperfine structure lines throughout the visible spectral range $3800 \lesssim \lambda \lesssim 8000$ \AA\@ (see Table \ref{tab:linelist} for absorption lines relevant for abundance analysis). 

The synthetic spectrum was Gaussian-smoothed using a standard deviation from the following equation:

\begin{equation} \label{eq:stdev-gaussian}
    \sigma_{\rm Gaussian} = \dfrac{\lambda_{\rm \hspace{.1em}central} }{2.35 R \Delta \lambda}
\end{equation} 

\noindent The central wavelength $\lambda_{\rm central}$ is unique for each spectral order in a particular exposure. For the order, we smoothed the relevant wavelength region of the synthetic spectrum with several resolving powers ranging from $20,000 \leq R \leq 50,000$. For each Gaussian-smoothed synthetic spectrum with its unique resolving power, we found a $\chi^2$ between the order and smoothed synthetic spectrum. The best resolving power was chosen to be the one with the lowest $\chi^2$, and we Gaussian-smoothed the full synthetic spectrum with this optimal resolving power. $\Delta \lambda = 0.01$ \AA \hspace{.1em} is the wavelength spacing set in {\fontfamily{qcr}\selectfont MOOG} to generate a synthetic spectrum.

For each exposure, we then divided each reduced echelle order by the smoothed synthetic spectrum in the particular spectral range of the order. This step essentially removes the absorption lines to give a continuum-only spectrum.  We then fit a third-order Chebyshev polynomial to the quotient spectrum using the function {\fontfamily{qcr}\selectfont fit\_generic\_continuum} from the Python {\fontfamily{qcr}\selectfont
specutils} package \citep{Earl2024}; default values in {\fontfamily{qcr}\selectfont fit\_generic\_continuum} were used. Finally, we divided the observed spectrum by the fitted quotient spectrum to normalize the continuum to unity. 

To co-add all orders to generate a single spectrum, we rebinned all continuum-normalized orders onto a common wavelength array and added the orders together via inverse variance weighting. The weights were determined from the uncertainties in each reduced echelle order dataset from  {\fontfamily{qcr}\selectfont MAKEE}\@. If a star had multiple exposures, we co-added all continuum-normalized exposures together using the same methodology to co-add the orders within a single exposure.

The S/N was computed using the difference between the observed spectrum and the model spectrum generated by {\fontfamily{qcr}\selectfont
MOOG} and dividing the model continuum by the median of this residual, i.e., $\frac{S}{N} = \frac{1}{\rm{median}([\text{cont-normed - model}])}$. \\

\subsection{Stellar Parameters}
\label{sec: stellar_params}

Abundances were measured from the HIRES echelle spectra by comparing them to synthetic spectra. Such an approach requires stellar parameters -- effective temperature $T_{\rm eff}$, surface gravity $\log(g)$, and microturbulence $\xi$ -- 
 which were derived using photometry from Gaia DR3 \citep{GAIA2023} for all stars. % except R9 and R21.
 Extinction corrections were applied in the following steps.  Using the Python interface {\fontfamily{qcr}\selectfont
dustmaps} \citep{Green2018}, we queried the dust reddening map from \cite{Schlegel1998}. The resulting $E(B-V)$ values are fairly uniform given the low spatial resolution of the \cite{Schlegel1998} data across M5, with an average $E(B-V) = 0.037$. The extinction was converted to the Gaia passbands $G, B_p, R_p$ using equation ($1$) of \cite{GAIA2018} and applied to the Gaia photometry.

After the extinction corrections, the stellar parameters for the model atmospheres were derived for all sample stars with readily identifiable photometry from Gaia. The parameter derivation is outlined in the rest of section \ref{sec: stellar_params}.  These stars have an assumed metallicity [Fe/H] = $-1.29$ \citep{Carretta2009a} and an $\alpha$ enhancement of [$\alpha$/Fe] = $0.4$ \citep{Ivans2001}. However, as \citet{Lai2011} stated, small changes in the $\alpha$ enhancement have an insignificant effect on the abundances measured. 

\begin{figure*} 
    \centering
    \includegraphics[width=\linewidth]{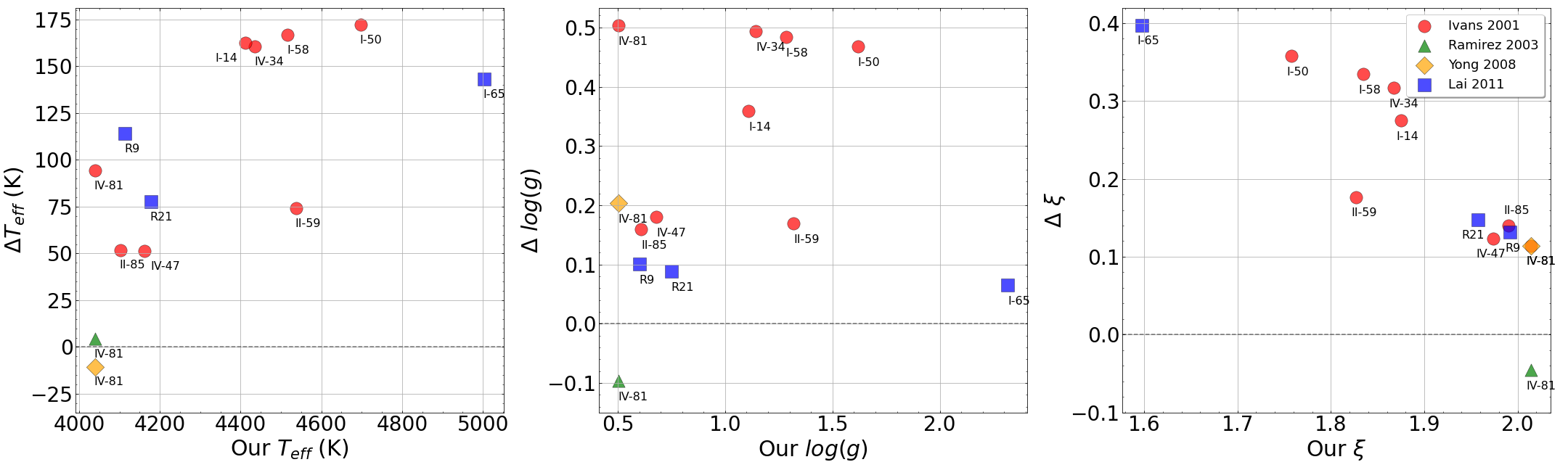}
    \caption{Comparison between our stellar parameters (see Table \ref{tab:stellar_data} for values) and literature data. }
    \label{fig:stellar_param_comp}
\end{figure*}

Our derived stellar parameters match well to the literature (\citealt{Ivans2001, Ramirez2003, Yong2008, Lai2011}) (see Fig.\ \ref{fig:stellar_param_comp}). The largest difference exists for the stars from \citet{Ivans2001} as they derived $T_{\rm eff}$ using spectroscopic methods, whereas our temperatures are photometrically derived (see \ref{sec: temperature} for more details). This difference is expected based on comparisons between spectroscopically and photometrically derived temperatures  \citep{Frebel2013}.

%Since the other two stars R9 and R21 were not readily identified with previous photometric surveys, we could not confidently associate the R.A.\ and Dec.\ for these stars in the KOA with the coordinates from Gaia. As a result, we could not find the appropriate Gaia DR3 identifier and thus did not have Gaia photometry for stellar parameter derivation. As a result, we instead utilized the stellar parameters for R9 and R21 reported in Table $3$ in \citet{Lai2011}.

\subsubsection{Temperature}
\label{sec: temperature}

While the effective temperature can be derived photometrically or spectroscopically, we computed all $T_{\rm eff}$ values using Gaia DR3 photometry \citep{GAIA2023}. Photometric temperatures are calibrated to physical temperature (from the infrared flux method) and are more reliable than spectroscopic temperatures, which in our study would be subject to all the shortcomings of the 1D, LTE code {\fontfamily{qcr}\selectfont
MOOG} \citep{Sneden2012}. These shortcomings include non-local thermodynamic equilibrium (NLTE) effects and an underestimation of red giant temperatures \citep{Frebel2013}. Avoiding spectroscopically derived temperature also ensures more accurate surface gravity. $T_{\rm eff}$ was computed via equation $(1)$ of \cite{Mucciarelli2021}, resulting in a temperature range of $4039 \leq T_{\rm eff} \leq 5003$.  

\subsubsection{Surface Gravity}

\label{sec: logg}

Surface gravity is computed from the Stefan--Boltzmann law:

\begin{equation} \label{eq:surface-grav}
    g = \frac{4 \pi GM \sigma_{SB} T_{\rm eff}^4}{L}
\end{equation}

\noindent Since all the sample stars are ancient, metal-poor RGBs, we assumed a mass $M = 0.75~M\textsubscript{\(\odot\)}$. We calculated the luminosity $L$ from the extinction-corrected Gaia $G_0$ magnitude, the distance 7.5~kpc \citep{Sandquist1996}, and applied bolometric
corrections (see Equation $(7)$ of \citealt{Andrae2018}). The $T_{\rm eff}$ in the equation described in section \ref{sec: temperature}.

\subsubsection{Microturbulence \texorpdfstring{$\xi$}{xi}}

Using the KOA data, there are two methods to compute the value of microturbulence $\xi$: minimizing the slope of abundance vs.\ line strength for absorption lines of the same species or via a parameterized fit, e.g., equation $(2)$ of \cite{Kirby2009}. Microturbulence was computed using this parameterized fit to avoid artificial systematic trends with evolutionary state, where $\xi$ for the bright (faint) giants would be fit preferentially to strong (weak) lines.

\subsection{Abundance Measurements}

 \label{sec:abun-measurements}
 
Radial velocity (RV) corrections, continuum refinement, and abundance measurements were performed with the  code Spectroscopy Made HardeR ({\fontfamily{qcr}\selectfont SMHR}) \citep{Casey2014}, which is a multi-purpose, comprehensive, high-resolution spectral analysis code for ascertaining chemical abundances. For RV measurements, each star was cross-correlated with a library spectrum ($6000 \leq \lambda \leq 6100$ \AA) of the reference RGB star HD 122563. After shifting the HIRES spectrum to the rest frame, pixels well above the continuum were removed with $0.2$-sigma clipping using the ``Normalization'' tab in {\fontfamily{qcr}\selectfont SMHR}\@.

%{\fontfamily{qcr}\selectfont SMHR} ran several {\fontfamily{qcr}\selectfont MOOG} syntheses for each \ion{Fe}{1} and \ion{Fe}{2} line in the line list of \cite{Ji2020} (see Table \ref{tab:linelist} for full line list) for each star. The $\log(\epsilon)$ abundances were compared to excitation potential $\chi$ and reduced equivalent width $\log({\rm EW} / \lambda)$.  Outliers were removed to reduce the slopes in $\log(\epsilon)$ plots. While stellar parameters can be derived from these slopes, we relied exclusively on the photometry-based stellar parameters computed in Section \ref{sec: stellar_params}.  

Once the spectra was adjusted and the stellar parameter values were set based on photometry (see Section \ref{sec: stellar_params}), we allowed {\fontfamily{qcr}\selectfont SMHR} to automatically compute abundances for all lines in our line list excluding the neutron-capture elements (see Table \ref{tab:linelist} for full line list \citep{Ji2020}). {\fontfamily{qcr}\selectfont SMHR} fit spectral lines with Gaussians to compute EWs and then ran several {\fontfamily{qcr}\selectfont MOOG} syntheses for each non-neutron capture element line in the line list for each star to compute an abundance. Although we did find abundances for several light elements, only Na and O are reported to differentiate the first and second stellar generations within M5. For these two elements, the lines used were in the redder wavelength region, meaning that line blending is minimal.  In particular, the Na lines are well separated from telluric contamination. 

The neutron-capture elements -- Ba, Nd, and Eu -- required special treatment. As blending and hyperfine structure splitting can contribute to the EWs of these neutron-capture element lines, we constructed individualized line lists with $10$~\AA\ ranges centered on each of the neutron-capture element lines in Table \ref{tab:linelist}.  We adopted $r$-process isotopic ratios from \cite{Sneden2008}. We set the [C/Fe] ratio using the parameterized fit in equation $(2)$ of \cite{Kirby2015}. Accounting for [C/Fe] had an insignificant effect on any measurement. All neutron-capture abundances were then computed line-by-line using the hyperfine structure (HFS) feature in {\fontfamily{qcr}\selectfont SMHR}\@. Table \ref{tab:stellar_data} shows the abundances of all relevant species in our study.

\begin{comment}
\st{The purpose of this study is to determine whether M5 has a systematic scatter in $r$-process abundances.  We are not searching for one or two outlier examples.  Therefore, outlier abundances were removed once all abundance analysis was completed. For each element, outliers were determined via the Inter-Quartile Range (IQR) method; only stars within the range $Q1 - 1.5\hspace{.1em} IQR \leq \log \hspace{.1em} \epsilon \leq Q3 + 1.5 \hspace{.1em} IQR$ were used to measure abundance dispersion. For Nd, the values of Q1 and Q3 are $0.326$ and $0.469$, respectively while for Eu, the values of Q1 and Q3 are $0.579$ and  $0.725$, respectively.}
\end{comment}

%Each reported elemental abundance has an associated uncertainty, which is the quadrature sum of the $\log\epsilon$ uncertainty reported by {\fontfamily{qcr}\selectfont SMHR} with the uncertainties on the abundance due to stellar parameter error propagation. The $\log\epsilon$ uncertainty from {\fontfamily{qcr}\selectfont SMHR} is the standard deviation for each species' abundance divided by the number of lines for each species---$\sigma / \sqrt{N}$---also known as the standard error (SE)\@. For some stars, only $1$ Eu or $1$ Nd line was present, so $N = 1$ for those particular species during our error analysis. In this case, the uncertainty $\sigma$ is only due to the error in the equivalent width analysis. 

Each reported elemental abundance error is computed via {\fontfamily{qcr}\selectfont SMHR} using a quadrature sum of the following: the abundance error due to the stellar parameters, the abundance error due to the covariance among the stellar parameters (Eq.\ B6 of \citealt{Ji2020}), and an abundance error due to systematic uncertainties (Eq.\ B14 of \citealt{Ji2020}).  The abundance uncertainties due to the stellar parameter errors were found in a two-step manner. First, each star's $b_p$ and $r_p$ errors from GAIA were propagated onto the effective temperature, surface gravity, and microturbulence parameters. Then, using the {\fontfamily{qcr}\selectfont SMHR} Python code {\fontfamily{qcr}\selectfont run\_errors}, we propagated the stellar parameter errors by systematically altering each parameter upwards and downwards by the relevant error value and running several abundance syntheses to c2apture the range of abundance changes due to changes in the parameter.

The {\fontfamily{qcr}\selectfont SMHR} Python code {\fontfamily{qcr}\selectfont make\_abund\_table} was then utilized to compute the error on the abundance due to stellar parameter covariance and also compute a systematic abundance uncertainty for each line analyzed. For abundances computed via spectral synthesis, i.e., the neutron-capture lines, an additional systematic uncertainty of $\sigma_{\rm sys} = 0.1$ was added to account for atomic data uncertainties. The output of {\fontfamily{qcr}\selectfont make\_abund\_table} is an abundance table with the quadrature sum of all three errors. The reported error is for a given line. For a particular species with multiple lines, the final abundance and error is the weighted sum of the individual lines. 

Of the 28 stars in our sample, $22$ were observed before the 2004 HIRES upgrade, thereby limiting the possible lines for which we could compute EWs and abundances for each of these ``pre-upgrade'' stars. Lines with $\lambda \lesssim 5200$~\AA\ were not observed in the pre-upgrade stars.

We report only abundances of the main $r$-process rare-earth elements Nd and Eu, along with Ba. This study focuses on the main $r$-process, rather than the weak/1st peak $r$-process elements (such as Sr). Unlike the common weak $r$-process, the main $r$-process only occurs within the most exotic environments, and out of the main $r$-process elements viewable in this wavelength region of visible spectra, only Nd and Eu were useful elements across the whole sample. The post-2004 upgrade stars, of which there are not many, do have lines for other rare-earth peak elements, but we do not report these abundances since our focus on dispersion in the $r$-process requires a large number of data points.

\begin{longtable}{lccccc}

\caption{Line list for neutron-capture species in this study, adopted from Table 3 of \citet{Ji2020}. Not every line was used in each star due to data quality or pre-2004 CCD limitations.
\label{tab:linelist}} \\

\hline
Species & Wavelength (\AA) & EP (eV) & $\log gf$ & Num. Stars \\
\hline
\ion{O}{1} & 6300.304 & 0.0 & -9.72 & 19 \\
\ion{O}{1} & 6363.776 & 0.02 & -10.19 & 18 \\
\ion{Na}{1} & 5682.633 & 2.102 & -0.71 & 28 \\
\ion{Na}{1} & 5688.203 & 2.104 & -0.41 & 28 \\
\ion{Na}{1} & 5889.951 & 0.0 & 0.11 & 28 \\
\ion{Na}{1} & 5895.924 & 0.0 & -0.19 & 9 \\
\ion{Ba}{2} & 4554.034 & 0.0 & 0.17 & 7 \\
\ion{Ba}{2} & 4934.1 & 0.0 & -0.157 & 5 \\
\ion{Ba}{2} & 5853.69 & 0.604 & -1.01 & 28 \\
\ion{Ba}{2} & 6141.73 & 0.704 & -0.077 & 28 \\
\ion{Ba}{2} & 6496.91 & 0.604 & -0.38 & 26 \\
\ion{Nd}{2} & 4109.45 & 0.32 & 0.35 & 2 \\
\ion{Nd}{2} & 4133.35 & 0.32 & -0.49 & 5 \\
\ion{Nd}{2} & 4358.167 & 0.32 & -0.16 & 3 \\
\ion{Nd}{2} & 4400.82 & 0.064 & -0.6 & 5 \\
\ion{Nd}{2} & 4446.387 & 0.204 & -0.35 & 1 \\
\ion{Nd}{2} & 4451.98 & 0.0 & -1.1 & 1 \\
\ion{Nd}{2} & 4462.98 & 0.559 & 0.04 & 2 \\
\ion{Nd}{2} & 4465.06 & 0.0 & -1.36 & 6 \\
\ion{Nd}{2} & 4501.81 & 0.204 & -0.69 & 7 \\
\ion{Nd}{2} & 4542.6 & 0.742 & -0.28 & 1 \\
\ion{Nd}{2} & 4645.76 & 0.559 & -0.76 & 7 \\
\ion{Nd}{2} & 4706.54 & 0.0 & -0.71 & 1 \\
\ion{Nd}{2} & 4820.34 & 0.204 & -0.92 & 7 \\
\ion{Nd}{2} & 4825.48 & 0.182 & -0.42 & 1 \\
\ion{Nd}{2} & 4902.04 & 0.064 & -1.34 & 1 \\
\ion{Nd}{2} & 4959.12 & 0.064 & -0.8 & 5 \\
\ion{Nd}{2} & 5092.79 & 0.38 & -0.61 & 3 \\
\ion{Nd}{2} & 5130.59 & 1.303 & 0.45 & 1 \\
\ion{Nd}{2} & 5234.19 & 0.55 & -0.51 & 6 \\
\ion{Nd}{2} & 5249.58 & 0.975 & 0.2 & 7 \\
\ion{Nd}{2} & 5255.51 & 0.204 & -0.67 & 3 \\
\ion{Nd}{2} & 5293.16 & 0.822 & 0.1 & 13 \\
\ion{Nd}{2} & 5319.81 & 0.55 & -0.14 & 16 \\
\ion{Eu}{2} & 4129.708 & 0.0 & 0.22 & 3 \\
\ion{Eu}{2} & 4205.026 & 0.0 & 0.21 & 1 \\
\ion{Eu}{2} & 4435.568 & 0.207 & -0.11 & 7 \\
\ion{Eu}{2} & 4522.573 & 0.207 & -0.67 & 1 \\
\ion{Eu}{2} & 6645.104 & 1.379 & 0.12 & 27 \\
\hline

\end{longtable}

\begin{comment}
THIS IS MADE 7-21-25 as the version of linelist with incorrect Nd count (issue was not flooring wavelenghts)
    \ion{Ba}{2} & 4934.1 & 0.0 & -0.157 & 5 \\
\ion{Ba}{2} & 5853.69 & 0.604 & -1.01 & 28 \\
\ion{Ba}{2} & 6141.73 & 0.704 & -0.077 & 28 \\
\ion{Nd}{2} & 4109.45 & 0.32 & 0.35 & 2 \\
\ion{Nd}{2} & 4358.167 & 0.32 & -0.16 & 1 \\
\ion{Nd}{2} & 4451.98 & 0.0 & -1.1 & 1 \\
\ion{Nd}{2} & 4462.98 & 0.559 & 0.04 & 2 \\
\ion{Nd}{2} & 4820.34 & 0.204 & -0.92 & 1 \\
\ion{Nd}{2} & 4825.48 & 0.182 & -0.42 & 1 \\
\ion{Nd}{2} & 5092.79 & 0.38 & -0.61 & 3 \\
\ion{Nd}{2} & 5130.59 & 1.303 & 0.45 & 1 \\
\ion{Nd}{2} & 5234.19 & 0.55 & -0.51 & 3 \\
\ion{Nd}{2} & 5249.58 & 0.975 & 0.2 & 7 \\
\ion{Nd}{2} & 5255.51 & 0.204 & -0.67 & 3 \\
\ion{Eu}{2} & 4205.026 & 0.0 & 0.21 & 1 \\
\ion{Eu}{2} & 4435.568 & 0.207 & -0.11 & 6 \\
\ion{Eu}{2} & 4522.573 & 0.207 & -0.67 & 1 \\
\ion{Eu}{2} & 6645.104 & 1.379 & 0.12 & 27 
\end{comment}

\section{Results}
\label{sec:results}

Most globular clusters display light element abundance dispersions such as the Na--O or Al--Mg anti-correlations \citep{Carretta2009}. These anti-correlations are the main evidence for the multiple-population/multiple-generation phenomenon. However, not many GCs are known to exhibit complex distributions of elements heavier than the iron peak. Focused studies on heavy element dispersions can hint at $r$-process production sites. We describe the light element and main $r$-process element distributions (with Nd and Eu) for M5 using our RGB sample.  

\subsection{Light Elements}

Our KOA sample reproduces the well-established \citep{Carretta2009, Lai2011} light element Na--O anti-correlation in M5. While Na--O anti-correlations have been found across most of the CMD---RGBs, AGBs, HBs, main sequence \citep{Gratton2019}---we focus on only one evolutionary state (RGBs) to minimize uncertainties arising from stellar models. 

\begin{figure}[h!]
    \centering
    \includegraphics[width=1\linewidth]{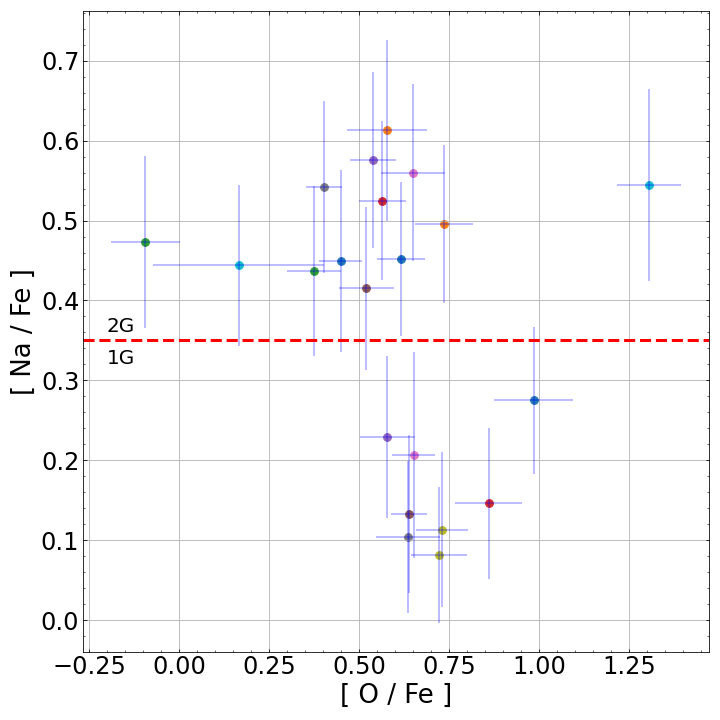}
    \caption{Na--O anti-correlation within the cluster sample. Based on this anti-correlation, we defined the first generation as stars with [Na/Fe] $\leq +0.351$ and the second generation as stars with [Na/Fe] $> +0.351$. This cutoff was computed using a K-means clustering with $N = 2$ groups to find the delineating abundance between the two groups. }
    \label{fig:light-element-anticorrelation}
    \label{fig:Na-O}
\end{figure}

Using the $21$ RGBs in Figure \ref{fig:Na-O}, we defined the separation between first- and second-generation stars to be [Na/Fe] = $0.351$ where the first (older) generation has [Na/Fe] $\leq 0.351$ and the second (younger) generation has [Na/Fe] $\geq 0.351$. Note that not all stars had spectra with measurable oxygen lines. 

For Na, the ISM Na contamination is well separated in velocity from stellar NaD\@.  Furthermore, the difference in abundance derived from the  NaD\@ lines and the weaker $5680$~\AA\ doublet has a median difference of just $+0.26$~dex. Within the $5680$~\AA\ doublet, the median Na abundance difference is just $+0.05$~dex.

NLTE corrections were not applied to our abundances. Based on NLTE corrections from \citet{Bergemann2021}, we found insignificantly small O NLTE corrections. For Na, we queried the NLTE corrections from \citet{Lind2011}. %Based on only the LTE abundances for Na $5682$~\AA\ line, the original [Na/Fe] value delineating the two generations in M5 was $+0.41$ dex, but the new [Na/Fe] value assuming an NLTE correction for Na $5682$~\AA\ is $+0.57$.
The median NLTE correction for the Na $5682$~\AA\ absorption line was $+0.16$ dex. Still, even though an NLTE correction shifted each star, there was no change in classifying stars into each generation, and thus any dispersion statistics were unaffected. Thus, since NLTE effects do not affect the classification by stellar generation, the reported abundances do not include NLTE corrections.

Note, as a consequence of using an RGB sample observed primarily before the 2004 HIRES upgrade, the Mg--Al anti-correlation was not used to define a cutoff because only $2$ stars from our sample included any prominent Al lines.

\subsection{Neutron-Capture Elements}
\label{sec:ncap}

Several neutron-capture abundances in M5 have been reported before in the literature \citep{Ivans2001,Ramirez2003, Yong2008, Lai2011, Roederer2011}.
This study focuses exclusively on Ba, Nd, and Eu as most of the sample stars only have spectra with these three neutron-capture element lines.  

\begin{figure*}
    \centering
    % Top Figure
    \begin{minipage}{\linewidth}
        \centering
        \includegraphics[width=.50\linewidth, height=9.5cm]{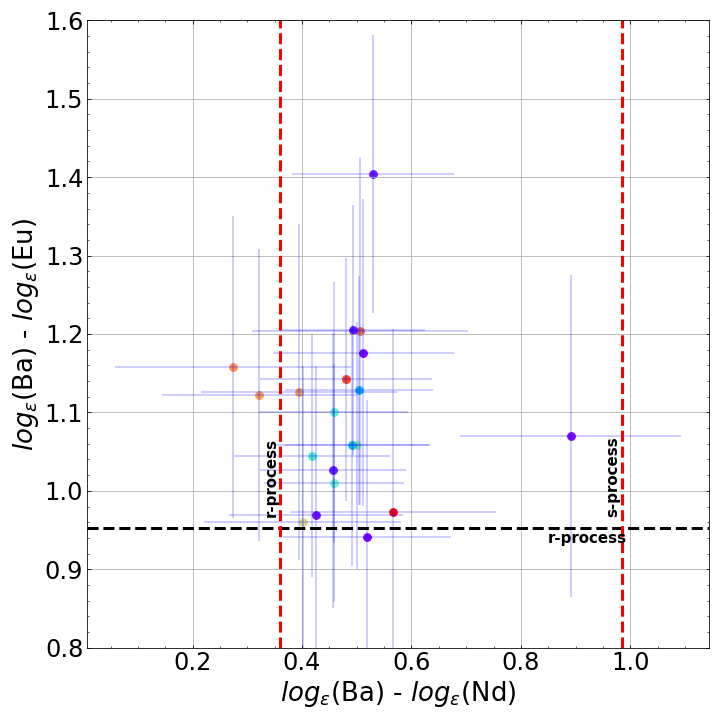}
        %\label{fig:s-r-ratio}
    \end{minipage}
    \vspace{1em} % Add vertical space between rows
    % Bottom Row
    \begin{minipage}{0.49\linewidth}
        \centering
        \includegraphics[width=\linewidth, height=9cm]{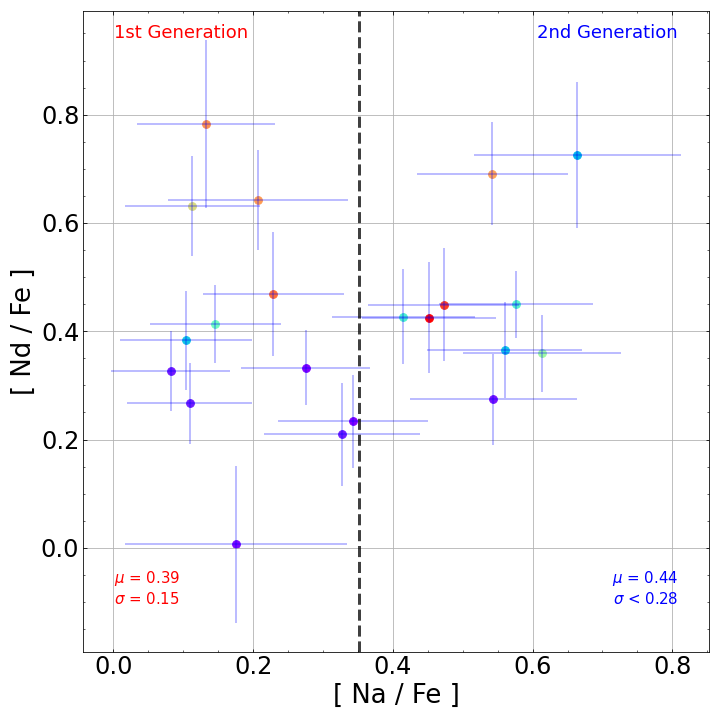}
        %\label{fig: Eu-dispersion}
    \end{minipage}
    \hfill
    \begin{minipage}{0.49\linewidth}
        \centering
        \includegraphics[width=\linewidth, height=9cm]{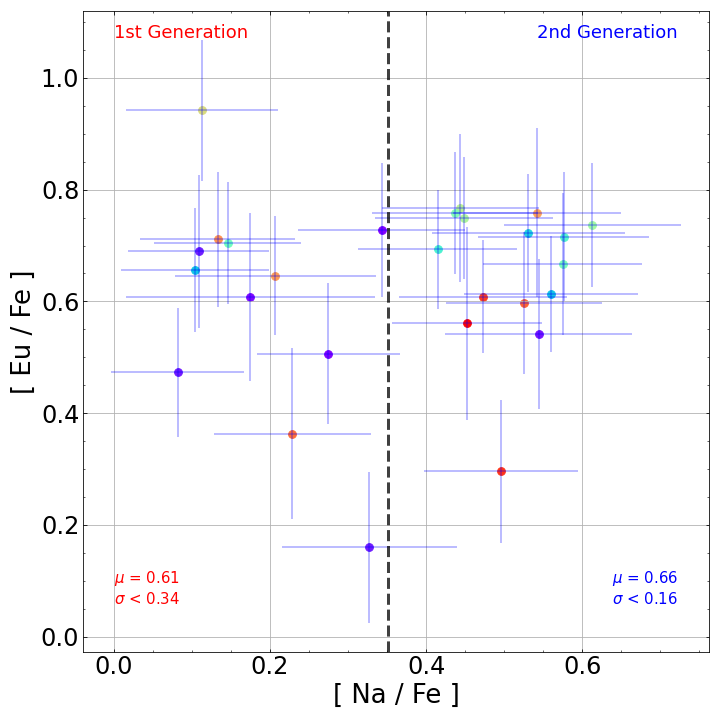}
        %\label{fig: Nd-dispersion}
    \end{minipage}
    % Main Caption
    \caption{(a) Comparison of RGB abundance ratios to the S.S.\ $s$- and $r$-process patterns \citep{Simmerer2004}. \textbf{The Nd and Eu abundances are nearly pure $r$-process. The $s$-process ratio between Ba and Eu is $\log_\epsilon ({\rm Ba})-\log_\epsilon({\rm Eu})=3.27$, indicating almost exclusively $r$-process nucleosythesis}.  (b) Neodymium dispersion as determined by the combination of Nd lines from Table \ref{tab:linelist} in each generation. (c) Europium dispersion based on the combination of Eu lines from Table \ref{tab:linelist} in each generation as defined by Fig. \ref{fig:light-element-anticorrelation}. }
    \label{fig:ncap-dispersion}
\end{figure*}

Eu and Nd, which are proxies for the main $r$-process in our study, can be formed either through the $s$-process or $r$-process, and observed abundances are often a combination of both neutron-capture processes. Since our study focuses on the origin of the main $r$-process, we utilized Solar System (S.S.) ratios from \cite{Simmerer2004} to distinguish between stars with purely $r$-process-derived abundances and those with $s$-process contributions. In the S.S., Ba is a primarily $s$-process element, presumed to have formed in AGB stars, while Eu and Nd are primarily $r$-process elements, with Nd having some non-negligible $s$-process contribution. Unlike in the S.S., the [Ba/Eu] and [Ba/Nd] ratios in M5 (Figure \ref{fig:ncap-dispersion}(a)) are consistent with the $r$-process. Thus, any neutron-capture abundance dispersion would be free of significant $s$-process contamination. This trend is demonstrated for all stars except for one Nd outlier (discussed in Section \ref{sec:outliers}).  

To isolate the intrinsic abundance dispersion, $\sigma_i$, in the cluster is difficult to separate from the abundance uncertainties, we employed an error-weighted dispersion study by maximizing the following \added{the negative log of the } Gaussian likelihood function:

\begin{equation}
-\ln(L) = \frac{\ln(2\pi)}{2} ~  + \frac{1}{2} \left[ \sum_{i=1}^{n} (\ln(\sigma^2 + \delta x_i^2) + \dfrac{(x_i - \mu)^2}{(\sigma^2 + \delta x_i ^2)}) \right ] 
\end{equation}

where $x_i, ~ \delta x_i$ represents an element's bracket abundance [X/Fe] and abundance uncertainty (see Sec.\ \ref{sec:abun-measurements}), respectively. $\mu$ is the average elemental abundance. The total likelihood $\prod_i L_i$ is maximized by following a procedure based on the library {\fontfamily{qcr}\selectfont astroML} \citep{Plas2014}, varying the mean $\mu$ between the minimum and maximum abundance and intrinsic dispersion $\sigma_i$ between $0 ~ - 3\sigma_w$, where $\sigma_w$ is the weighted standard deviation of the abundances based on the uncertainties $\delta x_i$. The resulting intrinsic dispersion and mean abundance for all three neutron-capture elements are reported in Table \ref{tab:statistics} for both generations and all stars in the starlist. 

%The significance of the resulting intrinsic dispersions can be interpreted via a reduced chi-squared value $\chi_{r, X}^2$ (root-mean-square normalized by uncertainty for element X). Here, the expected/model value for each star is the mean $\mu$ for each element. A $\chi_r^2$ chi-squared value less than $1$ indicates that the abundance dispersion across the whole sample is smaller than the measurement uncertainties, i.e., we overestimated the dispersion. A reduced chi-squared greater than $1$ implies some intrinsic dispersion within the sample, but since the confidence goes as the square root of the $\chi_r^2$, a value of $\chi_r^2 \gtrapprox 4$ ($95.5 \%$ confidence) would imply a strongly significant dispersion. The values computed for each element and generation are available in Table \ref{tab:statistics} for all sets of data.}

With low significance, we report the existence of main $r$-process dispersion within the globular cluster M5 based on Nd, specifically within the first generation but not within the second generation. For Nd, the $\sigma_i~(\rm{1G}) = 0.15$, which is different from zero at slightly more than $2\sigma$ significance.  The intrinsic dispersions for Ba and Eu in both generations and for Nd in 2G are consistent with zero within $2\sigma$ and are thus reported with upper limits.

%Nd is measured from five different absorption lines.  Different stars rely on a different combination of Nd lines.  Therefore, Nd is especially subject to systematic errors.  Therefore, our reporting of a Nd dispersion should be treated with caution.

%Thus, we corroborate the findings of M15 and M92 \citep{Kirby2023,Cabrera-Garcia2024}. Like M15 and M92, M5 exhibits a dispersion in the main $r$-process abundances in the first generation but little to no dispersion within the second. However, the results are tenuous due to the low number of elements and issues with both elements.

While the S/N for the Nd lines are high, this element was measured using 11 unique lines, most of which are not shared amongst the stars, i.e., 21 stars have Nd abundances reported from differing lines (see Table \ref{tab:linelist}). This inconsistency adds systematic error, and it may inflate the dispersion found for Nd. %On the other hand, while 27 out of the 28 stars had the Eu $\lambda6645$ \AA ~ line, the S/N in this region was occassionally low, resulting in potentially dubious abundance measurements.}

\newpage
\begin{longtable}{lcccccc}
\caption{Statistics for the [X/Fe] abundances, where X is a neutron-capture element. $\mu$ are reported for a $95\%$ confidence interval. Only Nd 1G and Nd all stars report lower and upper  $1 \sigma$ error bars for the dispersion. Other dispersion values report $2 \sigma$ upper limits. 
\label{tab:statistics}} \\
\hline
Element & $\sigma_i$ (1G) & $\sigma_i$ (2G) & $\sigma_i$ (All) & $\mu$ (1G) & $\mu$ (2G) & $\mu$ (All)  \\
\hline
\endhead
\hline
\endfoot

Ba & $< 0.16$ & $<0.22$ & $<0.13$ & $0.04_{-0.10}^{+0.10}$ & $0.05_{-0.09}^{+0.09}$ & $0.05_{-0.06}^{+0.06}$ \\

Nd & $0.15_{-0.07}^{+0.10}$ & $<0.28$ & $0.12_{-0.05}^{+0.07}$ & $0.39_{-0.14}^{+0.14}$ & $0.44_{-0.11}^{+0.11}$ & $0.41_{-0.09}^{+0.09}$\\

Eu & $<0.34$ & $<0.16$ & $<.18$ & $0.61_{-0.14}^{+0.14}$ & $0.66_{-0.08}^{+0.08}$ & $0.64_{-0.07}^{+0.07}$

\end{longtable}

%we report no strong or significant abundance dispersion within each generation for either Nd or Eu: $\chi_{r}^2 < 1.5$ for both populations (see Figure \ref{fig:ncap-dispersion}(b) and (c) for exact values).  
%We defined outliers using the Inter-Quartile Range (IQR) method, and III-66 is at the edge of the upper bound. If this star is removed from the calculation, then our result is insignificant with $\chi_{r} = .72$.

\subsection{Nd Outlier}
\label{sec:outliers}

Within our M5 sample, most stars exhibit similar [Nd/Fe] ratios. However, using the IQR method with the 1st and 3rd quartile, which have abundances of $0.326$ and $0.469$, respectively,  we identified $1$ outlier star: the Nd-depleted star, IV-3, in the first (low-Na) generation with [Nd/Fe] $= 0.01$. This is also apparent using the mean Nd abundance. Since the first generation has a mean $\mu(\text{Nd/Fe}) = 0.39$ and a standard deviation $\sigma(\text{[Nd/Fe]}) =0.15$, IV-3 is an Nd outlier. We compared the \ion{Nd}{2} $\lambda \, 5319$ absorption feature between IV-3 and other stars with similar stellar parameters %(see zoom-in plot of Figure \ref{fig:CMD_spectra} for the list of stars)
and confirmed a distinct equivalent width relative to the others. However, IV-3 is not an outlier for the Eu abundance. We also verified that IV-3 is a member of the M5 cluster based on radial velocity, proper motion, and coordinate measurements from Gaia DR3.  This star merits further scrutiny in a future work.

%The Nd abundance of III-66 is at the edge of the upper IQR bound. If this star is removed from the calculation, then the Nd dispersion is insignificant, with $\chi_{r} = .72$.

\section{Discussion}

Similar to the results reported by \cite{Roederer2011}, our study suggests M5 might exhibit dispersion in the main $r$-process elements. This result, however, is not as clear as the dispersions found the globular clusters M15, M92, and NGC 2298 \citep{Sneden2000,Kirby2023, Cabrera-Garcia2024, Bandyopadhyay2025}.

Based on the dispersions found within each generation of Nd, M5 might have a spread in abundances within the first generation but not the second. As a result, no main $r$-process event could have occurred between the formation of 1G and 2G within M5. Two pathways could explain this difference in abundance spreads between the generations. %One, the nebular gas that eventually formed the M5 globular cluster experienced a main $r$-process enriching event just before the formation of the globular cluster or during the formation of the 1G stars. In this case, the gas was polluted by these heavy metals asymmetrically, resulting in certain stars being polluted with more metals than others. The pollutants mixed homogeneously with the globular cluster nebular gas faster than the formation timescale of the second generation, i.e., approximately $30$ Myr. As a result, the second generation displays uniform main $r$-process abundances.

First, if the 1G gas was polluted by a star born concurrently with 1G, the main $r$-process source must have been prompt. There are only a few scenarios that could address such a prompt yet prolific event. Supernovae are low-mass, common, and short delay-time events that can produce $r$-process material. However, contemporary realistic 3D supernova simulations with neutrino transport find the electron fraction increases to values ineffective for the production of main $r$-process elements \citep{Wanajo2023}. Neutron star mergers have long been a proposed site of the $r$-process \citep{Lattimer1974}, and they are an observationally confirmed phenomenon \citep{Abbot2017}. However, their long delay-times of at least $\sim 30$ Myr in dynamical environments like GCs\citep{Kalogera2001} make it unlikely that a NSM resulting from massive stars that formed in the cluster exploded before the end of the first generation. Short delay-time NSMs in dynamical environments have been theorized \citep{Zevin2019}, but it is also possible that a very early NSM that exploded before M5 formed stars could have contributed $r$-process material to the nascent cluster. The remaining proposed sites that are also prompt, prolific, and rare include magneto-rotational supernovae \citep{Nishimura2015}, collapsars \citep{Siegel2019}, and magnetar flares \citep{Patel2025}. 

In the second pathway, the first generation formed due to the merging of various gas clouds, prompting globular cluster formation. In this scenario, each gas cloud consisted of unique main $r$-process abundances; the inhomogenous mixing of the gas during 1G formation results in a spread in Nd and Eu. The second generation forms similarly to the first pathway, i.e. due to even mixing of nebular gas.

\begin{comment}
To explain this lack of $r$-process dispersion,  we propose that the $r$-process elements were well mixed before the formation of M5, resulting in uniform Eu and Nd abundances. As the second generation of stars also exhibits the same mean main $r$-process abundance as the first generation, no main $r$-process occurred between the first and second generations, highlighting the rarity of a main $r$-process source.
\end{comment}

The Eu abundances found within M5 are consistent with those found in the Milky Way halo \citep{Sneden2008}, possibly suggesting that the two share a common source of the $r$-process. 
Given that most globular clusters and the halo share similar metallicities, it is possible that rare $r$-process events(s) polluted the MW halo environment and are responsible for pre-enriching the GCs. The dispersions found within M15, M92, NGC 2298, and potentially M5 might be exceptions to uniform $r$-process abundances associated with the halo and most GCs, meaning they experienced a distinct pathway for $r$-process enrichment.

The star IV-3 is a first-generation, Nd-deficient star. The existence of this star in our data does not rule out any particular type of $r$-process source or cluster formation scenario. The Eu abundance of this star is not particularly low.  Therefore, it is difficult to connect its abundances to a particular $r$-process source.  Our work does not fully explain this star's abundance pattern.

\subsection{Future Improvements For Detecting Abundance Dispersion}

Detecting heavy element dispersion in M5 using this KOA sample is prone to various errors. For one, this KOA sample consists of stars with a wide range of stellar parameters, thereby introducing uncertainties from model atmospheres. Also, only a few usable neutron-capture lines were present in the spectra as most stars were observed before the 2004 HIRES upgrade. Having only one Eu line for several stars limits the abundance analysis, especially if the Eu line is weak/noisy. 

Furthermore, M5 is a globular cluster with a metallicity [Fe/H] = $-1.29$ \citep{Harris1996}, which is about $10$ times more metal-rich than the previously confirmed clusters with $r$-process dispersion, M15, M92, and NGC 2298. As a result, the same linear dispersion in $r$-process abundances amongst the clusters would be present as a logarithmic [$r$/Fe] dispersion in M5 that is $10$ times smaller than the confirmed clusters. Given these issues, we can only conclude that dispersion was detected with low significance.

The simplest technique to study abundance dispersion for main $r$-process elements is to compare high-resolution spectra of stars with similar stellar parameters ($T_{\rm eff}$, $\log g$, $\xi$) at particular lines of interest. As an example, Figure \ref{fig:CMD_spectra} shows clear dispersion in the light element sodium for the hottest 7 stars in our sample without needing to consult a radiative transfer code. Analyzing stars with similar stellar parameters avoids many of the uncertainties of stellar abundance analysis, namely the assumptions used for the model atmosphere, oscillator strength uncertainties, and the assumption of LTE, all of which are utilized by {\fontfamily{qcr}\selectfont MOOG} to compute abundances. %Since M5 is more metal-rich, stacking spectra of similar stellar-parameter stars can also show weaker dispersion signals. 

In upcoming work, we will analyze stars in M5 in a tightly localized region of the CMD\@. This strategy will circumvent most systematic issues, increasing the detectability of $r$-process dispersion, if present. 

\vspace{-.4em}

%\section{Conclusion/Future Work}
%The lack of detectable $r$-process dispersion is not indicative of no dispersion. Detecting $r$-process dispersion within M5 is difficult since the  metallicity of M5 ($[Fe/H] = -1.29)$ is $\approx 10 x$ greater than M15 ($[Fe/H] = -2.37$ and M92 ($[Fe/H] = -2.31$) implies [r/Fe] is
%$\approx 10 x$ lower, assuming the pure abundance dispersion is the same between different metallicity GCs. Yadda yadda yadda.

\begin{acknowledgments}

We appreciate the referee's comments in refining this paper. We thank all the people who assisted with this project. This includes Shivani Shah and Kaitlin Webber for their continuous assistance with {\fontfamily{qcr}\selectfont SMHR} and Ian Roederer for his ideas and input on $r$-process dispersion. We also acknowledge the PIs whose data we utilized from the KOA: R.\ Kraft, J.\ Cohen, and M.\ Bolte. 
Portions of this text were written with the assistance of ChatGPT \citep{openai}, especially in regard to English usage and grammar.

We are thankful for the Arthur J. Schmitt Foundation's generous support during this project.

This research has made use of the Keck Observatory Archive (KOA), which is operated by the W.\ M.\ Keck Observatory and the NASA Exoplanet Science Institute (NExScI), under contract with the National Aeronautics and Space Administration. This research has made use of the SIMBAD database, operated at CDS, Strasbourg, France.

\end{acknowledgments}

\begin{contribution}
P.~Nalamwar performed the majority of the analysis and wrote the manuscript.  E.N.~Kirby provided advice and edited the manuscript.  A.~Cai wrote the first version of the code {\fontfamily{qcr}\selectfont StarSpec} that prepares the spectra for abundance analysis (Section ~\ref{sec: spectrum}).
\end{contribution}

\facilities{Keck: I (HIRES)}

\software{{\fontfamily{qcr}\selectfont MOOG}, {\fontfamily{qcr}\selectfont SMHR}, {\fontfamily{qcr}\selectfont ATLAS9}, {\fontfamily{qcr}\selectfont LineMake}}


\vspace{5mm}


\bibliography{References}{}
\bibliographystyle{aasjournal}



\end{document}